\documentstyle[aps,prl,graphicx,multicol]{revtex}

 \title{Electron spin decoherence  in 
quantum dots due to 
interaction with nuclei}

 \author{Alexander V. Khaetskii$^{1}$, Daniel Loss$^{1}$, and Leonid Glazman$^{2}$}
\address{$^{1}$
Department of Physics and Astronomy,
University of Basel,
Klingelbergstrasse 82,
CH-4056 Basel, Switzerland}

\address{$^{2}$
Theoretical Physics Institute, University of Minnesota, Minneapolis, MN
55455, USA.}

\begin{document}
\draft

\maketitle

\begin{abstract}

We study the decoherence  of a single electron spin  in an isolated quantum dot 
induced by  hyperfine interaction with nuclei for  times smaller than
the nuclear spin relaxation time. 
The decay is caused by the spatial variation of the electron envelope
wave function
within the dot, leading to a non-uniform hyperfine coupling.  
 We evaluate the spin correlation function with and without magnetic fields and
find 
that the decay of the spin
precession amplitude is not exponential but rather  power (inverse logarithm) 
law-like. For fully polarized nuclei we find an exact solution  and show that
the precession amplitude and the  decay behavior  can be tuned by the magnetic field.
   The corresponding  decay time is given by $\hbar N/A$, where 
$A$ is a hyperfine interaction constant and $N$  the number of  nuclei
inside the dot.  The amplitude of precession, reached as a result of the decay,
is finite.  
We show that there is a striking difference
between the 
decoherence time for a single dot and the dephasing time
for an ensemble of dots.
\end{abstract}

\pacs{PACS numbers: 85.35.Be; 71.21.La; 76.20.+q; 76.60.Es}

\begin{multicols}{2}

The spin dynamics of  electrons in semiconducting
nanostructures has become of central interest in recent years \cite{Wolf}.
The controlled manipulation of spin, and in particular of its phase, 
is the primary prerequisite needed for novel applications in conventional
computer hardware as well as in quantum information processing.
It is thus desirable to understand the  mechanisms
which limit the spin phase coherence of electrons, in particular in 
GaAs semiconductors, which have been shown to exhibit unusually
long spin decoherence times $T_2$ exceeding 100 ns \cite{Awschalom}.
Since in GaAs  each nucleus carries spin, the hyperfine
interaction between electron and nuclear spins is unavoidable, and it
is therefore important to understand its effect on the electron
spin dynamics. This is particularly so
for  electrons which are confined
to a closed system such as a quantum dot with a spin 1/2 ground state,
since, besides fundamental interest, these systems are promising candidates for 
scalable spin
qubits \cite{Loss98}. For recent work on spin relaxation (characterized by $T_1$ times)
in GaAs
nanostructures we refer to Refs.\cite{Khaetskii,Nazarov,Lyanda}.

Motivated by this we
investigate in the following the spin dynamics of a single electron
confined to a quantum
dot in the presence of nuclear spins. We treat  the case of unpolarized
nuclei perturbatively, while for the fully polarized case we present
an exact solution for the spin dynamics and show that the decay
is non-exponential and can be strongly influenced by external magnetic fields.
 We
use the term "decoherence" to describe the case with a single dot,  
and the term "dephasing" for an ensemble
of  dots \cite{ensemble}. 
The typical
fluctuating nuclear magnetic field  seen by the electron spin 
via the hyperfine
interaction   is of the order of\cite{Dyakonov}
$\sim A/(\sqrt{N} g \mu_B)$, with an associated electron precession
frequency  $\omega_N \simeq A/\sqrt{N}$,
where $A$ is a hyperfine constant, $g$  the
electron g-factor, and  $\mu_B$  the Bohr magneton.
 For a  typical dot size the electron wave function covers approximately $N=10^5$
nuclei, then this field is of the order of 100 Gauss in a GaAs quantum dot.
The nuclei in turn interact with each other via dipolar
interaction, which does not conserve the total nuclear spin
and thus leads to a change of a given nuclear spin configuration within the time
$T_{n2} \approx 10^{-4} s $, which is just the period of precession of a nuclear
spin in the local magnetic field  generated by its neighbours.

We note that  there are two    different regimes of interest,  
depending on the parameter $\omega_N \tau_c$, where
$\tau_c$ is the correlation time of the  nuclear field.
  The  simplest case is given by the perturbative regime
$\omega_N \tau_c \ll 1$,  
characterized by dynamical narrowing: 
different random nuclear configurations  change quickly in time, and, as a
result, the spin dynamics is diffusive with a
dephasing time $ \simeq 1/ (\omega_N^2 \tau_c)$.  
A more difficult situation arises when $\omega_N \tau_c \gg 1$, requiring
a nonperturbative approach. It is this regime which we will
consider in this paper, i.e.  the electron is localized in a quantum dot, and  the
correlation time is due to the internal nuclear spin dynamics, i.e., $\tau_c=T_{n2}$,
giving $\omega_N \tau_c=10^4$. 
Next, we need to address the important issue of averaging over different nuclear 
spin configurations in a single dot. Without internal nuclear 
spin dynamics, i.e. $T_{n2}\rightarrow  \infty$, no averaging is indicated. However,
each flip-flop process  (due to hyperfine interaction)
creates a different
nuclear configuration, and because of the spatial variation of the hyperfine
coupling constants inside the dot, this
leads  to a different value of the  nuclear  field seen by the
electron spin and thus to its decoherence. Below we will find
that this decoherence is non-exponential, but still we can indicate a
characteristic time given by $(A/\hbar N)^{-1}$\cite{ensemble}. Moreover, we shall find
that $T_{n2}\gg (A/\hbar N)^{-1}$, and thus
 still no averaging over the nuclear configurations is indicated (and
dipolar interactions will be neglected henceforth).
To underline the importance of this point, we will  contrast below
the unaveraged correlator with its average.

{\it Unpolarized nuclei}. 
We consider a single electron confined to a quantum dot whose spin ${\bf S}$
couples to an external magnetic field ${\bf B}$ and to
nuclear spins $\{{\bf I}^i\}$ via hyperfine contact interaction, described
by  the Hamiltonian 
\begin{equation}
\hat {\cal H}= g\mu_B {\bf S} \cdot {\bf B} + {\bf S} \cdot {\bf h}_N\, , \/
 \, {\bf h}_N =  \sum_i A_i {\bf I}^i\, , 
\end{equation}   
where ${\bf h}_N/(g\mu_B)$ is the nuclear field.
The
coupling constant with the i-th nucleus, $A_i =A v_0 | \Psi({\bf r}_i) |
^2$, contains 
the  electron  envelope
wave function $\Psi({\bf r}_i)$ at the nuclear site ${\bf r}_i$, and $v_0$ is the
volume of the crystal cell.  We start with the case ${\bf B}=0$,
and for simplicity we consider nuclear spin 1/2.
   Neglecting dipolar interactions between the
nuclei, we can  consider only  some particular nuclear configuration,
described in the $\hat I_z^i$ eigenbasis as
$| \{I_z^i\}>$, with $I_z^i=\pm 1/2$.
Moreover, we
assume an unpolarized configuration with a typical net nuclear magnetic
field $A/(\sqrt{N}g\mu_B)$, being much less than 
$ A/(g\mu_B)$ (fully polarized case).
 We study now the decay  of the electron spin from its initial ($t=0$)
$\hat S_z$-eigenstate $| \Uparrow >$.
For this we evaluate the correlator  
$C_n(t)=<n | \delta \hat  S_z(t) \hat S_z | n>$,
where $\delta \hat S_z(t) =  \hat S_z(t)- \hat S_z$, and
$\hat S_z(t) = \exp(it \hat {\cal H}) \hat S_z \exp(-it \hat {\cal H})$.
Since at $t=0$ the total (electron and nuclear)
state $|n>=|\Uparrow, \{I_z^i\}>$ is an eigenstate of $ \hat {\cal H}_0 = \hat S_z \hat 
h_{Nz}$ (with eigenergy $\epsilon_n$),   we can expand in the perturbation    $\hat
V= (1/2)(\hat S_+ \hat h_{N-} + 
\hat S_- \hat h_{N+})$ (with $\hat {\cal H}= \hat {\cal H}_0 + \hat V$).
Going over to the interaction picture,  we obtain in leading order
\begin{equation} 
C_n(t) =  \sum_k \frac{|V_{nk}|^2}{\omega_{nk}^2} (
\cos (\omega_{nk} t)  -1 )\, ,
\label{4}
\end{equation} 
where $V_{nk}=<n|\hat V|k>$ is the matrix element 
between initial state
$n=\Uparrow, \{..., I^k_z = -1/2,... \}$  and intermediate state $k= \Downarrow, 
\{...., I^k_z = +1/2,... \} $,  and $\omega_{nk}= \epsilon_n -\epsilon_k$.
Using $|V_{nk}|^2  = 
A_k^2 <n| 1/2- \hat I^k_z | n>/4$, and $\omega_{nk} = (h_z)_n + A_k/2$, where
$(h_z)_n = <n | \hat h_{Nz} | n>$, we obtain for the typical nuclear
configuration, for which $(h_z)_n^2 \simeq \omega_N^2 \gg A_k^2$, 
\begin{eqnarray}  
&C_n(t)&
\simeq  \gamma
\left [I_0- I_1 (\tau)\cos
((h_z)_n t)  + I_2(\tau) \sin ((h_z)_n t) \right], \nonumber \\   
 &I_{i}(\tau)&
=\int_{-\infty}^{+\infty} dz \chi_0^4(z)F_{i}(\tau
\chi_0^2(z)), \,\, i=0,1,2,
\label{cor1}
\end{eqnarray} 
where $\gamma=-A^2/(8\pi N (h_z)_n^2)$, 
$F_0=1/2$, $ F_1(\eta)= \sin \eta/{\eta} + (\cos
\eta -1)/{\eta^2}$, and $  F_2(\eta)= \sin \eta/{\eta^2} - \cos \eta/{\eta}$.
Here, $N = a_z a^2/v_0 \gg 1$ is the number of  nuclei inside the
dot, and
$\tau =  A t /2\pi N$.  We have replaced the sums over $k$ (which run over the
entire space)  by integrals and  used that
$| \Psi(\rho,z)|^2 = (1/\pi a^2 a_z) \exp(- \rho^2/a^2) \chi_0^2(z)$.
Here 
$a, a_z$ are the dot sizes in the lateral and transverse (perpendicular to the
2D plane) directions, resp.,
and  the transverse wave
function  is normalized, i.e. 
$\int_{-\infty}^{+\infty} dz
\chi_0^2(z)=1$.  For any analytic function  with  expansion
$\chi_0^2(z)= \chi_0^2(0)-z^2(\chi_0^2)''/2$ near its maximum, we have
$I_{1,2}(\tau \gg 1)= \pm (\chi_0^2(0)/\tau^{3/2}) 
\sqrt{\pi/(\chi_0^2)''} [\sin (\tau \chi_0^2(0))
\mp \cos(\tau \chi_0^2(0))]$.  
 Thus, $C_n(t) \propto 1/\tau^{3/2}$, i.e. the spin decay follows a
{\em power law} for times
$\tau \gg 1$, i.e. $t \gg (A/N)^{-1}$.    Note that for the
typical nuclear configuration the quantity 
$A^2/N(h_z)_n^2 $ is of  order unity, thus the part of the 
 electron spin state which decays is of the order of the initial value.
Hence the same holds for the spin part which survives at $\tau \gg 1$. 
For the fully polarized nuclear state the result (\ref{cor1}) should be
multiplied by 2, and $(h_z)_n$ should be replaced by $A/2$. Moreover,
in the presence of
a large Zeeman field, $\epsilon_z= g\mu_B B_z
\gg \omega_N$, we should substitute $\epsilon_z$ for $(h_z)_n$. 
 Thus, the asymptotic behaviour of $C_n(t)$  is not
changed, the only difference being that 
the decaying part of the initial spin state  is small now, being of the order of
$(\omega_N/\epsilon_z)^2 \ll 1$.

We note  that $C_n(t)$ in (\ref{4}) is 
quasiperiodic in $t$, and, thus, it will decay  only up to the Poincare recurrence
time $\tau_P$.
 This time
can be found from the condition that the terms  omitted when converting  sums
to  integrals  become comparable with the integral itself.
This will happen at $\tau
\simeq N$, giving $\tau_P=0.1s - 1 s$. 

In  next order, $\hat V^4$, we face the problem of 
"resonances", i.e. the  corrections
will contain  zero denominators. This gives rise to linearly growing terms
$\propto\omega_N t$, even for $t
\ll (A/N)^{-1}$. In higher order  the degree of the divergence
will increase.     This means that the decay law we found can, in principle,
change after proper resummation, 
because 
no   small expansion parameter  exists, which, strictly speaking, would justify a perturbative
approach. Still, the result found in lowest order remains qualitatively correct in that it
shows that a non-uniform hyperfine coupling leads to a non-exponential decay of the
spin. This conclusion is confirmed  by an exactly solvable case to which we turn
next.

 {\it Polarized nuclei. Exactly solvable case.}  In this section we
consider the exactly solvable case where the initial
 nuclear spin configuration is fully  polarized. We also allow for a magnetic field
but neglect its effect on the nuclear spins.
With the initial wave function 
$\Psi_0 = |\Downarrow; \uparrow, \uparrow, \uparrow,...>$ we can construct the {\em exact}
wave function of the system for $t>0$, 
\begin{equation}
\Psi(t)= \alpha(t) \Psi_0 + \sum_k \beta_k(t) |\Uparrow; \uparrow, \uparrow,
\downarrow_k,\uparrow,...>,
\label{function}
\end{equation} with normalization $|\alpha(t)|^2 + \sum_k
|\beta_k(t)|^2 =1$, and we assume that $\alpha(t=0^+) =1, \/ \alpha(t<0) =0 $.
Then, inserting $\Psi(t)$ into the Schr\"odinger equation we obtain
\begin{eqnarray}
 i\frac{d\alpha(t)}{dt} &=& -\frac{1}{4} A
\alpha(t) + \sum_k \frac{A_k}{2}\beta_k(t)  -\frac{\epsilon_z \alpha(t)}{2},
\nonumber
\\  i\frac{d\beta_l(t)}{dt} &=& (\frac{A}{4}-\frac{A_l}{2} )\beta_l(t)  +
\frac{A_l}{2} \alpha(t)+\frac{\epsilon_z \beta_l(t)}{2},
  \label{system}
\end{eqnarray} where $A = \sum_k A_k $. 
 Laplace transforming (\ref{system}), we obtain 
\begin{equation}
\alpha(t)= \frac{1}{2\pi i} \int_{\Gamma} 
\frac{d \omega\, i \exp[(\omega -iA'/4 )t] }
{i\omega+ \epsilon_z + \pi N i\omega  \int
dz\ln (1-
\frac{iA \chi_0^2(z)}{2\pi N \omega})}\/,
\label{Zeeman}
\end{equation} where $A'= A + 2\epsilon_z$.  We have  replaced the sum 
$\sum_k \frac{A_k^2}{i\omega - A'/4 + A_k/2}$ over the  $xy$-plane   
 by an integral and  calculated it  using 
$|\Psi({\bf r}_k)|^2$ given above.  As usual, the integration contour $\Gamma$
in
Eq.(\ref{Zeeman}) is the vertical line in the complex $\omega$-plane so that
all 
singularities of the integrand  lie to its left.  These singularities are: 
two branch points ($\omega=0, \/
\omega_0=iA 
\chi_0^2(0)/2\pi N$),  and first order poles which lie on the imaginary axis
($\omega=iv$).  For  $\epsilon_z>0$ there is  one pole, while for 
 $\epsilon_z<0$ there  are two poles, and  for $\epsilon_z=0$ there is one 
first order pole at $\omega_1 \approx iA/2 + iA \int dz \chi_0^4(z) /4\pi N$. 
For the contribution from the branch cut between $\omega =0$ and $\omega =
\omega_0$ we obtain 
\begin{eqnarray}
&\tilde \alpha(t)& = \frac{e^{-iA't/4}}{\pi N} 
\int_0^1 {d\kappa  2z_0 \kappa e^{i\tau' \kappa}}  
\Big \{[\kappa \int dz \ln|1 - \frac{\chi_0^2(z)}{\chi_0^2(0) \kappa}| \nonumber \\
&+&  
\kappa/\pi N - 2\epsilon_z/A\chi_0^2(0)]^2 + (2\pi z_0)^2\kappa^2\Big \}^{-1},
\label{arbitr}
\end{eqnarray}  
where $\tau'= \tau \chi_0^2(0)$, and  $z_0=z_0(\kappa)$ is defined through
$\chi_0^2(z_0) = \chi_0^2(0) \kappa$. 
Let us consider first the case $\epsilon_z=0$.  The asymptotic behavior
 of the integral (\ref{arbitr}) for
$\tau \gg 1$ is determined by  $\kappa \ll 1$. For example, for
$\chi_0^2(z)/\chi_0^2(0)= \exp(-z^2)$ we find $\tilde\alpha \propto 1/\ln^{3/2}\tau $. 
  Thus, the decay  of $|\alpha(t)|$ starts at $\tau >1$,  i.e. at $t>
A^{-1}N$, as in the unpolarized case.  Note that the magnitude of
$\tilde \alpha$ is of  order  $1/N$, thus the decaying part of the initial
spin state has this smallness as well, in contrast to 
the unpolarized case above where
this part is of order unity\cite{BLD}.  For large Zeeman field ($|\epsilon_z| \gg
A$) and for $\tau \gg 1$, the main contribution in (\ref{arbitr}) is given for  $\kappa
\rightarrow 1$. Expanding
$\chi_0^2(z)$ for small $z$ (see above), we obtain $z_0^2= 2\chi_0^2(0)
(1-\kappa)/(\chi_0^2)''$. Then from Eq.(\ref{arbitr}) we have for
$|\epsilon_z| \gg A$
 \begin{equation}
\tilde \alpha(\tau \gg 1)= \frac{ -e^{i\tau'-iA't/4 }}{4\sqrt{\pi} N} 
 \frac{\chi_0^2(0)}{\sqrt{(\chi_0^2)''}}\frac{A^2}{\epsilon_z^2} 
\frac{(1+i)}{\tau^{3/2}}.
\label{arb}
\end{equation} From this we find then that the 
correlator $C_0(t)=-<\Psi_0| \delta \hat S_z(t)
\hat S_z|\Psi_0> =(1 - |\alpha(t)|^2)/2 $
agrees with the perturbative result (\ref{cor1}) for the fully polarized state,
i.e.  $ C_0(t)\propto 1/t^{3/2}$. This agreement is to be expected, since for
large Zeeman field, the perturbative treatment with a small parameter
$A/|\epsilon_z| \ll 1$
 is meaningful\cite{anisotropy}.
However, at zero Zeeman field, when the system cannot be treated
perturbatively, we find  $C_0(t)\propto 1/\ln^{3/2}t$, and the agreement
with (\ref{cor1}) breaks down.  Nevertheless, the characteristic time scale 
for the onset of the non-exponential decay is  the same for all cases and given by
$(A/N)^{-1}$. 

There are several interesting features which we can observe for the fully
polarized state. In an external Zeeman field, the effective gap seen by the
electron spin  is $A'/2= A/2 + \epsilon_z$. Thus,  when
$\epsilon_z$ is made negative this gap decreases and even vanishes
at $ |\epsilon_z|= A/2$.  {}From Eq.(\ref{Zeeman}) we find that 
the two poles are symmetric in
this case, and the system resonates between the two frequencies $\omega_{\pm} = \pm i
A(\int  \chi_0^4(z)dz)^{1/2}/\sqrt{8\pi N}$.  Note that the 
residual gap is of  order  $A/\sqrt{N}$ (and not $A/N$, as one might
naively expect).    Near this Zeeman field we have
$|\alpha(t)|^2=\cos^2(\omega_{+}t)$ (up to small corrections of order $1/N$),
and, as a result, $|\alpha|^2$ averaged over time  is 1/2, i.e. the
up and down states of the electron spin are strongly coupled via the nuclei. In contrast,
outside this resonance regime  the value  of $|\alpha|^2$ is
close to 1 (again with small $1/N$ corrections), i.e. 
$<\hat S_z(t)>=1/2-|\alpha|^2$ is close to
-1/2 at any time. The width of the resonance is $\sim A/\sqrt{N}$, i.e. small compared to
the initial gap $A/2$.    We note that this behavior represents
periodic (Rabi) oscillations with a single well-defined frequency and
is not related  to decoherence. [The latter is described by the branch cut 
contribution $\tilde \alpha$ which  remains small (order 1/N) even near 
the resonance.] This abrupt change  in the 
amplitude of oscillations of $<\hat S_z(t)>$ (when changing $\epsilon_z$ in
a narrow interval around $A/2$) can be used for an experimental detection of the fully
polarized state. Note that the weight of the upper pole alone (i.e. that which exists at
$\epsilon_z=0$) also  drops abruptly from a value close to 1 to a value much smaller
than 1 in the same narrow interval, which can be
experimentally checked by Fourier analysis.  Another special value of Zeeman
field corresponds to the case when the  upper pole is close
to $\omega_0$  ($\kappa
=1$)-- the upper  edge of the branch cut.  This occurs (see Eq.(\ref{Zeeman})) at
the  critical value
$\epsilon_z^{\star}=bA/2<0$, where $b= \chi_0^2(0)\int dz \ln
|1-\chi_0^2(z)/\chi_0^2(0)| < -1$ is a non-universal number which depends on the
dot shape.    Since at finite Zeeman field the asymptotics in $t$ is
determined by $\kappa$'s close to 1,  we see from Eq.(\ref{arbitr}) that
for $\epsilon \approx \epsilon_z^{\star}$ the asymptotics  changes
abruptly. Indeed, for $((\epsilon_z-\epsilon_z^{\star})/A)^2 \ll 1$,  we find
$\tilde\alpha \propto 1/\sqrt{\tau}$, for 
$ 1\ll \tau \ll  ((\epsilon_z-\epsilon_z^{\star})/A)^{-2}$, and $\tilde\alpha
\propto
1/\tau^{3/2}$, for
$\tau \gg ((\epsilon_z-\epsilon_z^{\star})/A)^{-2}$. Thus,  when approaching the
critical Zeeman field $\epsilon_z^{\star}$ there is a {\em slow down} 
of the asymptotics
from 
$1/\tau^{3/2}$ to
$1/\tau^{1/2}$.  It is interesting that this slow down is related to a strong
modulation of the density of states (DOS) of the excitations within the continuum band
(branch cut) near its edge when $\epsilon_z \rightarrow \epsilon_z^{\star}$. 
In the subspace of none or one
 nuclear spin flipped (see
Eq.(\ref{function})), the DOS becomes  $\nu(u)= Im [G_0(u) +
d/du
\ln D(u)]$, where $u=i\omega$, $G_0(u)= \sum_k 1/(u + A_k/2)$ is the "unperturbed
Green's function", and $D(u)$ is the denominator of $\alpha(\omega)$, see Eq.(\ref{Zeeman}).
One can then show that for $\epsilon_z \rightarrow \epsilon_z^{\star}$
(i.e. the upper pole
approaches the continuum  edge), the DOS  develops a square
root singularity: $\nu(u) \propto 1/\sqrt{\omega_0 - u}$. Simultaneously,  the weight
of the upper pole vanishes linearly in $\epsilon_z$ as
$\epsilon_z^{\star} -\epsilon_z\rightarrow 0$.   

Finally, the nuclear state is characterized by $\beta_k(t)$, which allows for similar
evaluation as $\alpha$. Here we just note that its branch cut part,
$\tilde \beta_k(t)$, is nonmonotonic 
in time, particularly pronounced at $\epsilon_z \rightarrow \epsilon_z^{\star}$: 
First $\tilde \beta_k(t)$ grows like $ \sqrt\tau $, until $\tau$
reaches
$\sim 1/(1-a_k)\gg 1$, and
then  it decays like $ 1/(\sqrt{\tau}(1-a_k))$, with $a_k=A_k/A_0
\rightarrow 1$. Thus,  $\beta_k$ is  maximal  for  $A_k$
close to $A_0$, i.e. the nuclei near the dot-center are 
affected most by the hyperfine
interaction with the electron spin.

{\it Averaging over nuclear  configurations}.  We have seen that the  decay
of  $C_n(t)$ occurs in the time interval $ N/A \ll t \ll
N^2/A$, with $N/A \simeq 10^{-6} s$ in GaAs dots. On the other hand,
the electron spin precesses in the net nuclear field (see Eq. (\ref{cor1}))
with the
characteristic period 
$(h_z)_n^{-1} \simeq \omega_N^{-1} 
\simeq 10^{-8} \div 10^{-9} s$.
Thus, $\omega_N^{-1}\ll N/A$, and we see that
the electron spin undergoes many  precessions   in a given
nuclear field before decoherence sets in due to the non-uniform hyperfine couplings
$A_k$.  
This  behavior changes dramatically when we average over  nuclear
configurations \cite{ensemble}.
For that purpose we consider high temperatures, $k_BT \gg \hbar \omega_N$,
and average $C_n(t)$ in Eq.(\ref{4}) over all
 nuclear configurations, i.e. $C(t)=\sum_n C_n(t)/\sum_n$. We then find
\begin{equation} 
C(t)=
\sum_k \frac{-A_k^2}{8} \int_0^t dt_1\int_0^t   dt_2
\prod_{i\neq k}\cos[\frac{A_i}{2}(t_1-t_2)]. 
\label{aver}
\end{equation} 
For $\tau << 1$, we get $\prod_{i\neq
k}\cos(A_it/2)= \exp[- NC (At/2\pi N)^2 ]$, where 
$C= \pi\int dz \chi_0^4(z)/4$.  Thus, the averaged spin correlator $C(t)$
(\ref{aver}) is of  order 
$- \int_0^{\omega_N t} dx \Phi (x)$, with 
$\Phi$ being the error function.  Thus, $C(t)$ 
grows without bound as $ \omega_N t$
for $ \omega_N t \gg 1$ (the condition $\tau \ll 1$ can still be 
satisfied). Consequently, the perturbative approach breaks  down  even
in leading order in $\hat V$ (we recall that {\em without} averaging the 
divergences  occur in all higher but not in lowest order).
To treat this case properly, we need a non-perturbative approach.
For that purpose we  calculate now the correlator $C(t)$
exactly by treating the nuclear field  purely classically, i.e. as a c-number.  Then we
obtain,  
\begin{equation}
 C_n(t) = - \frac{h_{N\perp}^2}{4 h_N^2} (1- \cos h_N t), 
\label{classic}
\end{equation} 
where $h_N = \sqrt{h_{Nz}^2 + h_{N\perp}^2}$ is the nuclear field, with $ \/
h_{N\perp}^2= h_{Nx}^2 + h_{Ny}^2$. The value of  $h_N$  corresponds to a given
nuclear configuration $n$. To make contact with the perturbation procedure which we used
before in the quantum case we go to the regime $h_{N\perp}^2 \ll
h_{Nz}^2$,  where $h_N$ can be replaced by  $h_{Nz}$ in Eq.(\ref{classic}).
Then we average the resulting expression  $(h_{N\perp}^2/ h_{Nz}^2) (1- \cos
h_{Nz} t)$  over a Gaussian distribution for $h_N$, i.e. over $P(h_N) \propto
\exp(- 3h_N^2/2\omega_N^2)$.  The result becomes  proportional to
$\int_{0}^{+\infty} dz \exp(-z^2/2) (1- \cos(\gamma z))/z^2 \propto
\int_0^{\gamma} dx
\Phi(x)$, where $\gamma= \omega_N t/\sqrt{3} $. Thus, we see that we obtain exactly the
same functional form as before from Eq.(\ref{aver}) with the same
divergencies in t. This reassures us that the  treatment of the nuclear
field as a classical field is not essential. On the other hand, the same Gaussian
averaging procedure can now be applied  to the non-perturbative form
Eq.(\ref{classic}). Defining $C_{cl}(t)=\int dh_N P(h_N) C_n(t)$, we obtain 
\begin{equation}
C_{cl}(t)= -\frac{1}{6} [1+ ( \frac{\omega_N^2
t^2}{3} -1)\, e^{-\omega_N^2 t^2/6}].
\label{exact}
\end{equation} 
Thus we get rapid (Gaussian)
 decay  of the correlator for $t\gg \omega_N^{-1}$, giving the dephasing time
$\omega_N^{-1}=\sqrt{N}/A$.
This means that $<\hat S_z(t) S_z>$ saturates at  1/3 of its
initial value of $1/4$. Finally, 
it seems likely that for the case of nuclear quantum  spins 
a non-perturbative treatment of the averaged  correlator $C(t)$
 will lead to a similar rapid time decay as found for the classical case in Eq.
(\ref{exact}).

In conclusion, we have studied  the  spin decoherence of an electron
confined to a single
quantum dot in the presence of hyperfine interaction with  nuclear spins. 
  The decoherence is due to a non-uniform coupling of the electron spin to 
nuclei located at different sites. The decoherence time is given by 
$\hbar N/A$ and is of the order of several $\mu s$. It is shown that in a weak
external Zeeman field the perturbative treatment of the electron spin
decoherence is impossible. Moreover, the decay of the electron spin
correlator in time does not have an exponential character, instead it is given by a power
or inverse logarithm law.     
We have shown that there is a strong  difference between
the  decoherence time for a single dot, 
$\hbar N/A$, and the dephasing time for an ensemble of dots, $\hbar\sqrt{N}/A$. 

We acknowledge  support from the 
Swiss NSF, DARPA and ARO. Part of this work
was performed at the Aspen Center of Physics and at the ITP, UC Santa
Barbara.

\end{multicols}


\begin{references}




\bibitem{Wolf} S.~A. Wolf  {\it et al.}, 
Science {\bf 294}, 1488 (2001).

\bibitem{Awschalom}
J. M. Kikkawa, D. D. Awschalom,
Phys.\ Rev.\ Lett.\ {\bf 80}, 4313 (1998).


\bibitem{Loss98}
D. Loss,  D. P. DiVincenzo,
Phys.\ Rev.\ A {\bf 57}, 120 (1998).


\bibitem{Khaetskii} A.V. Khaetskii, Yu.V. Nazarov, Phys. Rev. B {\bf 61}, 12639 (2000);
{\bf 64}, 125316 (2001).

\bibitem{Nazarov} S.I. Erlingsson, Yu.V. Nazarov, V.I. Fal'ko, Phys. Rev. B {\bf 64},
195306 (2001).

\bibitem{Lyanda} Y.B. Lyanda-Geller, I.L. Aleiner, B.L. Altshuler, cond-mat/0112013.



\bibitem{ensemble} We note that in cases with exponential decay the spin decoherence
(dephasing) time is usually denoted by $T_2$ ($T_2^*$) (e.g. in the Bloch equations).
E.g., in Ref.\cite{Awschalom} it is
$T_2^{\star}$ which is measured (due to an ensemble average over many spins). In
general, $T_2^{\star} < T_2$. 
Since the spin decay found in the present work turns out to be non-exponential, we will
not use this notation  (i.e. $T_2$ or $T_2^*$) to avoid confusion.



\bibitem{Dyakonov} 
M.I. Dyakonov and V.I. Perel, in  {\em Optical Orientation}, North-Holland, Amsterdam,
p. 11 (1984).


\bibitem{BLD}
G. Burkard, D. Loss, and D. P. DiVincenzo,
Phys.\ Rev.\ B {\bf 59}, 2070 (1999).


\bibitem{anisotropy}  The same is true for any model with a small expansion
parameter. E.g., for a system with 
anisotropy, where 
the hyperfine constants in perpendicular and transverse
directions are different, i.e. $A_i^z =A_z v_0 | \Psi({\bf r}_i) | ^2 \neq
A_i^{\perp} =A_{\perp} v_0 | \Psi({\bf r}_i) | ^2$, one obtains again
a change in the asymptotics from $1/\ln^{3/2}\tau$ (zero
anisotropy)  to $1/\tau^{3/2 }$ (large anisotropy). The anisotropy plays the role of
$\epsilon_z$.


\end{references}
\end{document}